\documentclass[journal]{IEEEtai}
\usepackage{amsmath,amsfonts}
\usepackage{algorithmic}
\usepackage{multicol}
\usepackage{algorithm}
\usepackage{array}
\usepackage[caption=false,font=normalsize,labelfont=sf,textfont=sf]{subfig}
\usepackage{textcomp}
\usepackage{url}
\usepackage{verbatim}
\usepackage{multirow}

\usepackage{cite}

\usepackage{textcomp}
\usepackage[table]{xcolor}

\usepackage[colorlinks,urlcolor=blue,linkcolor=blue,citecolor=blue]{hyperref}

\usepackage{color,array}

\usepackage{graphicx}

\begin{document}

\title{Biaxialformer: Leveraging Channel Independence and Inter-Channel Correlations in EEG Signal Decoding for Predicting Neurological Outcomes}

\author{Naimahmed Nesaragi, Hemin Ali Qadir, Per Steiner Halvorsen, and Ilangko Balasingham, \IEEEmembership{Member, IEEE}
\thanks{This work was supported in part by the Health South East
Authority in Norway, Helse Sør-Øst RHF (HSØ: New Realtime Decision Support during Blood Loss using Machine
Learning on Vital Signs) under Grant No. 19/00264–202, and
Prosjektnummer 2020079.}
\thanks{N.Nesaragi and H. A. Qadir are with  the Intervention Centre,
Oslo University Hospital, 0372 Oslo, Norway (e-mail:
naimahmed.nesaragi@gmail.com and hemina.qadir@gmail.com).} \thanks{P.S. Halvorsen is with Intervention Centre, Oslo University Hospital 0372, Oslo Norway, and also with Institute of Clinical Medicine, Faculty of Medicine, University of Oslo, Oslo, Norway
(e-mail: p.s.halvorsen@medisin.uio.no).}
\thanks{I. Balasingham is with the Intervention Centre, Oslo University Hospital
0372, Oslo Norway, and also with the Department of Electronic Systems,
Norwegian University of Science and Technology, 7491 Trondheim, Norway
(e-mail: ilangko.balasingham@medisin.uio.no).}}

\markboth{PREPRINT}
{First A. Author \MakeLowercase{\textit{et al.}}: Bare Demo of IEEEtai.cls for IEEE Journals of IEEE Transactions on Artificial Intelligence}

\maketitle

\begin{abstract}
Accurate decoding of EEG signals requires comprehensive modeling of both temporal dynamics within individual channels and spatial dependencies across channels. While Transformer-based models utilizing channel-independence (CI) strategies have demonstrated strong performance in various time series tasks, they often overlook the inter-channel correlations that are critical in multivariate EEG signals. This omission can lead to information degradation and reduced prediction accuracy, particularly in complex tasks such as neurological outcome prediction. To address these challenges, we propose \textit{Biaxialformer}, characterized by a meticulously engineered two-stage attention-based framework. This model independently captures both sequence-specific (temporal) and channel-specific (spatial) EEG information, promoting synergy and mutual reinforcement across channels without sacrificing CI.
By employing joint learning of positional encodings, \textit{Biaxialformer} preserves both temporal and spatial relationships in EEG data, mitigating the inter-channel correlation forgetting problem common in traditional CI models. Additionally, a tokenization module with variable receptive fields %adjusts kernel sizes and strides to 
balance the extraction of fine-grained, localized features and broader temporal dependencies. To enhance spatial feature extraction, we leverage bipolar EEG signals, which capture inter-hemispheric brain interactions, a critical but often overlooked aspect in EEG analysis. Our study broadens the use of Transformer-based models by addressing the challenge of predicting neurological outcomes in comatose patients. Using the multicenter I-CARE data from five hospitals, we validate the robustness and generalizability of \textit{Biaxialformer} with an average AUC 0.7688, AUPRC 0.8643, and $F_1$ 0.6518 in a cross-hospital scenario. 
\end{abstract}

\begin{IEEEImpStatement}
Decisions about continued treatment for comatose patients hinge on uncertain predictions of brain recovery, leaving families and clinicians in a difficult position. This work delivers a reliable AI‑based forecast of recovery chances by analyzing routine EEGs, consistently across multiple hospitals. This clarity can guide doctors toward personalized treatment plans, reduce the performance of invasive or costly procedures with little benefit, and give families timely, trustworthy information when weighing care options. By translating complex EEG signals into actionable insights, this approach promises to improve patient care, ease emotional burdens on families, and foster smarter resource allocation in critical care.
\end{IEEEImpStatement}


\section{Introduction}
\label{sec:introduction}
\IEEEPARstart{A}{nnually}, over 6 million cardiac arrests (CAs) occur worldwide, with survival rates ranging from 1\% to 10\% \cite{graham2015strategies}. After resuscitation from CA, the majority of patients are comatose and treated at intensive care units (ICUs) \cite{rundgren2010continuous}. Assessment of neurological outcomes in comatose CA survivors due to post-anoxic coma remains a persistent challenge in clinical practice \cite{rossetti2016neurological, perkins2021brain}. Electroencephalography (EEG) is one of the most available tools and the popular modality used for assessing unconscious patients, owing to its non-invasive nature, ease of use, and high temporal resolution \cite{amorim2023international}. EEG is inherently multivariate, meaning it comprises multiple time series, each representing the electrical activity recorded from a different electrode placed on the scalp. These time series capture rich information about brain function encoded in both intra-channel features within an electrode (e.g., frequency variations, power in specific bands, burst suppression patterns) and inter-channel features (e.g., correlations or differences in activity between electrodes, phase coherence). The interplay between these features holds crucial insights into brain states and intentions. However, the relative importance of these features can vary. %depending on the specific EEG classification task. 
For instance, sleep stage classification might prioritize intra-channel signatures of EEG signals to particular frequency bands based on the targeted sleep
stage \cite{boostani2017comparative}. Conversely, inter-channel EEG features are heavily relied in motor imagery classification for identifying differences in activity between channels at certain time points, such as event-related desynchronization/synchronization \cite{nam2011movement}.

In the context of neurological outcome prediction after CA, the challenge is further compounded by the complexity of brain dynamics during coma \cite{zheng2021predicting}. Here, both temporal dynamics within channels and spatial relationships across channels are critical for assessing brain function. For eg., burst suppression patterns within a channel may indicate severe brain injury leading towards poor prognosis, while inter-channel coherence may reveal preserved network connectivity, which is predictive of good prognosis \cite{sandroni2022prediction,rossetti2017electroencephalography,bongiovanni2020standardized,ruijter2019early}. This intricate balance between intra- and inter-channel information makes the underlying task more challenging. Moreover, the variability across patients adds another layer of complexity. EEG patterns in comatose patients can vary widely depending on individual factors, including the underlying cause of CA, duration of hypoxia, comorbidities, and post-resuscitation care. This results in a wide range of intra- and inter-channel feature variability, making it difficult to generalize across patients. Though multi-channel EEG streams aid in reducing the subjectivity of prognostic evaluation to some extent, a diverse variation still exists among intra- and inter-channel features. So, selecting the most suitable feature information becomes crucial. Several prognostic features eg., burst suppression, and nonreactive EEG patterns have already been identified based on the outcome of interest. However manual interpretable quantification of multivariate EEG streams is a laborious task
that demands advanced clinical and neuro-physiological expertise, limiting the accessibility of EEG-informed prognostication. Hence, automating EEG interpretation can overcome these barriers, enhancing accessibility and diagnostic accuracy in clinical settings.
 
Deep learning (DL) techniques have recently been increasingly applied to multivariate time series (MTS) EEG data, offering a promising solution to the challenges of Brain-Computer Interface analysis. 
%However, these state-of-the-art advancements have particularly underscored the impact of Transformer-based models on general MTS analysis.
%Transformer-based models have demonstrated improved predictive performance by refining the attention mechanism and have presented an intriguing avenue for further exploration of the multi-channel integration of the brain \cite{shi2022multimodal}. 
Based on modeling the variable dependencies, transformer-based models in general MTS can be broadly bifurcated into \textit{channel-independence (CI)} and \textit{channel-dependence (CD),} also called channel-mixing models \cite{han2024capacity,wang2023dance,han2024mcformer,wang2024card}.  Although \textit{CI} methods like PatchTST \cite{nie2022time} that focus on intra-channel processing typically claim to yield better results, they overlook the potential complex interrelationships between different channels of MTS data. Conversely, \textit{CD} methods like Crossformer \cite{zhang2023crossformer}, and iTransformer\cite{liu2023itransformer} could offer improvements by leveraging inter-channel correlations, but may lead to inadequate extraction of mutual information, and in some cases, potentially introduce noises \cite{wang2023dance}. Nonetheless, these studies argue that the integration of uncorrelated (cross-variable) information in the \textit{CD} strategy could decrease the potential improvement in MTS model performance. Based on the considerations outlined above, it is quite evident that the Transformer-based models still has substantial promise in multivariate EEG analysis, particularly if advancements are made in extracting meaningful information from inter- and intra-channel data. To substantiate this claim, we introduce a bi-axes transformer-based network called \textit{Biaxialformer}- that effectively leverages cross-variable information across inter-channels (electrodes) while adequately extracting temporal information of intra-channel (time-steps) simultaneously. \textit{Biaxialformer} introduces a sequence-channel-aligned two-stage attention structure that allows it to capture both temporal correlations among EEG sequences and dynamical dependence among multiple electrodes over time. To efficiently utilize a two-stage attention structure, we first design a joint learning of positional encodings to enhance the Transformer’s ability to model the spatiotemporal dynamics of EEG signals. The joint learning by positional encoding ensures that the model captures the spatiotemporal structure of the EEG data, which is crucial for tasks that require an understanding of both temporal patterns (e.g., rhythms, spikes) and spatial correlations (e.g., between neighboring electrodes). In this study, we hypothesize that a two-stage multi-head attention mechanism (MHA) can enhance the predictive power of multichannel EEG data to classify comatose patients with good or poor neurological outcomes. Specifically, we propose that the fusion of intra- and inter-channel attention mechanisms can effectively discern attention patterns within individual EEG sequences and cross-attention patterns among multiple EEG channels.
We summarize our key contributions as follows:
\begin{itemize}
%\vspace{-0.5em}
    \item We propose a two-stage MHA Transformer-based model that enhances both intra- and inter-channel features, promoting effective \textit{CI} by independently modeling each channel's temporal sequence simultaneously leveraging cross-variable information across electrodes.

    \item Our model incorporates joint learning of temporal and spatial positional encodings to address inter-channel correlation forgetting, a common issue in \textit{CI} models.
    
    \item We design a token fusion module with adjustable receptive fields, optimizing the balance between localized features and broader temporal dependencies.
    
    %\item This study addresses the critical need to extend the evaluation of transformer-based networks on EEG signals beyond the motor-imagery classification tasks such as emotion recognition and sleep-stage classification by evaluating the predictive power of multichannel EEG data for neurological outcomes in comatose patients post-CA.

    %\item  We adapt multi-center data (from 5 hospitals) in a cross-subject, cross-cohort scenario, addressing the limitations of small datasets, and utilize bipolar EEG signals to capture critical inter-hemispheric interactions.
    
\end{itemize}
\vspace{-1em}

\section{Realted Work}
\subsection{CI and CD modeling  and Transformer-based models for EEG decoding}

Many existing transformer-based time series prediction models have modeled global dependencies well. In the case of time series transformers, the longer the lookback window, the lower the performance, and the limitation is that computation increases explosively \cite{liu2023itransformer}. A vanilla transformer architecture embeds each time step across multivariate time-series as a temporal token leading to a mixed variate representation \cite{liu2023itransformer}. Unified embedding by each temporal token may fail to learn a variate-centric representation of multiple series or create a meaningless attention map. In other words, existing time series transformer-based models struggle to establish multivariate correlation.% as shown (we need a figure here)

PatchTST\cite{nie2022time} architecture splits the time series data into patches, allowing it to maintain \textit{CI} between different variables in the data. It divides lengthy time series into manageable \textit{patches} enabling efficient processing and capturing fine-grained patterns. \textit{CI} in PatchTST refers to the idea that instead of processing all channels together, it analyzes them individually, unlocking unique insights within each data stream.
However, while this approach excels at isolating and focusing on intra-channel features, it can struggle with inter-channel dependencies, potentially missing important long-range interactions between different channels. In other words,  \textit{CI} offered by PatchTST is a double-edged sword. This channel isolation can limit the model's ability to capture interactions between channels in MTS, leading to a phenomenon known as channel mixing.

Further, iTransformer\cite{liu2023itransformer} implementation tries to mitigate this problem of channel mixing where instead of several channels being embedded at once for each specific time-point, iTransformer embeds each univariate channel as an individual token. This leads to an advantage of variate-unmixed representation where each channel's contribution is modeled independently, allowing the model to avoid interference between less relevant channels. However, this strategy reduces the overall number of tokens available for each channel, and while it enables greater \textit{CI}, it may limit the model’s ability to capture long-range dependencies across channels due to its larger receptive field. iTransformer offers a compromise, enabling some \textit{CI} through masking, while still allowing interaction modeling via inter-channel relationships.

This variation in feature importance results in the challenge of effectively capturing these diverse intra- and inter-channel features across various EEG coding tasks, so selecting the most suitable model architecture becomes crucial. In summary, Patch-based transformer architectures like PatchTST excel at capturing intra-channel features due to their strong \textit{CI}, making it ideal for EEG tasks where individual channel processing is crucial, such as sleep stage classification. However, such architectures may struggle with capturing long-range inter-channel dependencies needed for tasks like motor imagery classification. Conversely, transformer architectures like iTransformer effectively model inter-channel interactions for tasks involving long-range dependencies but may face challenges in isolating the independent contributions of each channel due to the increased potential for interference across channels.

%EEG-based Transformer models have been used for emotionrecognition, classification of imagined speech, and sleep stageclassification [39]–[43].  critical need to extend the evaluation of transformer networks on EEG signals beyond the

%Another drawback of these studies is that they use within-subject training by combined EEG datasets and this approach has limited adaptability and robustness for different individuals [46]. than small data scenarios like within-subject and within-session applications, limiting the transformer model’srobustness

However, Transformer-based models in general have
demonstrated improved predictive performance by refining the
attention mechanism and have presented an intriguing avenue
for further exploration of the multi-channel integration of
the brain \cite{shi2022multimodal}. Few Transformer-based schemes \cite{song2022eeg,lee2022eeg,wang2022transformers,kim2022deep,kim2024towards} have garnered attention and brought
into EEG decoding applications but are confined mostly to motor-imagery classification tasks such as emotion recognition, and classification of sleep stages and imagined speech. Most of these studies are limited to the usage of  within-subject and within--
session applications, limiting the transformer model’s robustness for different individuals \cite{xie2022transformer,huang2022eeg}. Hence, there is a critical need to extend the evaluation of transformer-based networks on EEG decoding beyond these aforementioned applications.
\vspace{-1.009em}

%Here we need to write what we are proposed to solve these issues, and how we differ from other available models. 
\subsection{ Automated schemes for predicting neurological outcomes}
%\begin{list}{}{}
%\item{EEG-conformer} 
%\item{DFformer}
%\end{list}
The recent modest literature on predicting neurological outcomes using EEG recordings stems from the PhysioNet 2023 Challenge \cite{reyna2023predicting,amorim2023international} that aimed at the development
of various automated schemes focused on enhancing true positive rate (TPR) for
predicting a poor outcome given a false
positive rate (FPR) $\leq$ 0.05 at 72 hours
after the return of spontaneous circulation (ROSC). Most of these
state-of-art submissions\cite{reyna2023predicting,zabihi2023hyperensemble,yang2023model} rely on machine-learning (ML) methods using extraction of hand-crafted EEG
features like statistical, power in frequency bands, entropy values, etc., from multiple domains or transfer learning using pre-trained models \cite{kim2023predicting},
as it is more challenging to extract the required EEG feature representation
for the desired type of neurological outcome. Further, ML studies mostly predict outcomes based on EEG patterns observed within a specific time window, without utilizing the temporal evolution of the EEG data \cite{zheng2021predicting1}. 

Contemporary research on the prediction of
neurological outcomes in comatose patients
has employed various DL techniques to gain insight into EEG data by sequence modeling, and has yielded satisfactory
predictive performances \cite{tjepkema2019outcome,jonas2019eeg,pham2022outcome,zubler2023deep}. However, with hindsight, providing a straightforward
comparison among these studies
is tedious since, the
context or focus of interest in the predicted outcome is different, and variability
exists in the definitions of prediction windows \cite{zheng2021predicting}. Regardless of faithful results obtained by these DL models for the prediction of neurological outcomes, the existing
literature is naïve towards the noteworthy application of the transformer-based model \cite{chen2023electroencephalogram,jain2024emergence} that can capture the spatiotemporal structure of the sparse MTS EEG data, crucial for tasks that require understanding both temporal patterns among EEG sequences and spatial correlations between neighboring electrodes.

\section{Materials}

\subsection{Study Population}
The data considered under this study originates from the recent PhysioNet/Computing in Cardiology Challenge 2023 \cite{reyna2023predicting,amorim2023international}. This data is retrieved from ICU patients of seven academic hospitals in the U.S. and Europe led by investigators in the International Cardiac Arrest REsearch consortium (I-CARE). The goal is to employ EEG recordings to predict neurological outcomes among comatose patients after CA. Neurological function was determined using the Cerebral Performance Category (CPC) scale. CPC is an ordinal scale ranging from 1 to 5 with “Good outcome”: CPC = 1 or 2,
“Poor outcome”: CPC = 3, 4, or 5.  

This multi-center databases has adult patients with in-hospital or out-of-hospital CA who had ROSC but remained comatose. Each patient's brain activity was monitored with a 19-channel continuous EEG. Monitoring is typically started within hours of CA and continues for several hours to days depending on the patient’s condition, so recording start time and duration varied from patient to patient. The databases includes EEG data obtained up to 72 hours from ROSC. Challenge organizers obtained the consent of approval from the appropriate institutional review boards to collect data \cite{amorim2023international,reyna2023predicting}.
The data were split into training, validation, and test sets. The training set comprises of 607 patients (approximately 60\%), while 10\% of patients in the validation set, and 30\% in the test set.  %The training, validation, and test sets were chosen to approximately preserve the univariate distributions of each variable provided in the data. 
The proposed study has used the publicly shared dataset of 607 patients from five hospital centers (A, B, D, E, F) for the experiments in leave-one-out center format. Detailed statistics are mentioned in Table \ref{tab:samplecount}.
%\vspace{-1em}
\begin{table}[h]
\centering
\caption{\label{tab:samplecount} Distribution of neurological outcomes in publically shared training cohort}

\begin{tabular}{|l|l|l|l|l|l|}
\hline
\rowcolor{lightgray} \begin{tabular}[c]{@{}l@{}} Hospital\\  code\end{tabular} & \begin{tabular}[c]{@{}l@{}} Good\\Outcomes \end{tabular} & \begin{tabular}[c]{@{}l@{}}Poor\\Outcomes\end{tabular} & \begin{tabular}[c]{@{}l@{}}Total\end{tabular} & \begin{tabular}[c]{@{}l@{}}Study\\ Dataset \\ (\%)\end{tabular} & \begin{tabular}[c]{@{}l@{}}   Poor \\Outcomes\\(\%)\end{tabular}  \\ \hline
A                                                        & 122                                                                   & 139  & 261  & 43.00                                                                   & 53.26                                                                \\ \hline
B                                                        & 34                                                                   & 86  & 120  & 19.77                                                                   & 72.00                                                            \\ \hline
D                                                        & 27                                                                  & 56  & 83 & 13.67                                                                   & 67.47                                                                \\ \hline
E                                                        & 15                                                                   & 59    & 74  & 12.19                                                                   & 79.73                \\ \hline
F                                                        & 27                                                                   & 42 & 69   & 11.37                                                                   & 60.87                                                                 \\ \hline
Total                                                       & 225                                                                   & 382 &607 & - & -                                                               \\ \hline
\end{tabular}
\end{table}
\vspace{-2.5em}
\subsection{Pre-Processing}
The data pre-processing pipeline includes the following sequence of operations: filtering, re-sampling, rescaling, bipolar conversion, and finally, segmentation of EEG recordings. At the onset, the entire EEG data is filtered using a Butterworth band pass filter with cut-off frequencies of 0.5 Hz and 35 Hz to remove baseline wander and high-frequency noises from the EEG signals. Next, the filtered signals are examined in terms of sampling frequency, and all the EEG recordings are re-sampled to 100 Hz to maintain uniformity w.r.t sampling frequencies. The re-sampled signals are then re-scaled using Min-max standardization. Next, re-scaled signals are converted to bipolar representations and finally segmented. The bipolar conversion, subtracting EEG signals from adjacent scalp electrodes, is crucial in EEG signal processing \cite{yao2019reference}. It reduces noise and artifacts, enhances spatial resolution by focusing on localized brain activity, minimizes volume conduction effects, and aids comparisons to baseline states. It is valuable in clinical and neuroscience research and provides a cleaner and more accurate brain activity representation \cite{zaveri2006use}. This data pre-processing pipeline yields 18 bipolar channels from every hour of EEG recordings.

Further, in our study, we employ a systematic approach to process  $n$-minute EEG segments into subseries-level patches by the tokenization module (see subsection \ref{feature encoder}). We adopt the idea of patching\cite{nie2022time} since patches can better capture
local information and also encompass richer dependencies
between EEG channels and sequences.

\noindent Let $\mathbf{X}$  denote a multivariate time-series EEG dataset with $m$ samples of $n$-minute segments given as,  
\begin{equation}
\label{eqn1}
\mathbf{X}=\left\{\mathbf{\underset{1} {x}^{(t,k)}}, \mathbf{\underset{2} {x}^{(t,k)}}, \ldots, \mathbf{\underset{m} {x}^{(t,k)}} \right\}
\end{equation}

%$\mathbf{X}=\left\{\mathbf{x}_1, \ldots, \mathbf{x}_L\right\} \in \mathbb{R}^{B\times C \times L}$, where $B$ indicates the batch size, $C$ represents the number of channels in the EEG array, and $L$ represents the length of the channel. 

\noindent where, ${\mathbf{\underset{i} {x}^{(t,k)}} \in \mathbb{R}^{ C\times L}},  t=0,1,\ldots,L ; k=0,1,\ldots,C$.

\noindent $C$ denotes the number of EEG channels and $L$ denotes the length in samples of each EEG sequence corresponding to $n$-minute multivariate EEG array segment. $L$ in samples is  computed as  $L = n \times f_{s} \times 60$, where $f_{s}$ is the sampling frequency.

\noindent Let $\mathbf{\underset{i} {x}^{(k)}} \in \mathbb{R}^C$ represent the values of the various channels at the $t$-th time point and let $\mathbf{\underset{i} {x}^{(t)}} \in \mathbb{R}^L$ denote the input univariate EEG time series of the $k$-th channel. \noindent Let $\mathbf{Y}$  denote label set for the  corresponding EEG dataset $\mathbf{X}$ with $m$ samples given as,  
\begin{equation}
\label{eqn1}
\mathbf{Y}=\left\{\mathbf{\underset{1} {y}}, \mathbf{\underset{2} {y}}, \ldots, \mathbf{\underset{m} {y}} \right\}
\end{equation}
\noindent where, ${\mathbf{\underset{i} {y} \in  \left\{0,1 \right\}}}$, 0- Good,  1- Poor.

\section{Methods} 
\label{sec:methods}
\begin{figure*}[!t]
\centering
\includegraphics[trim=0 8 0 0, clip, scale=0.63]{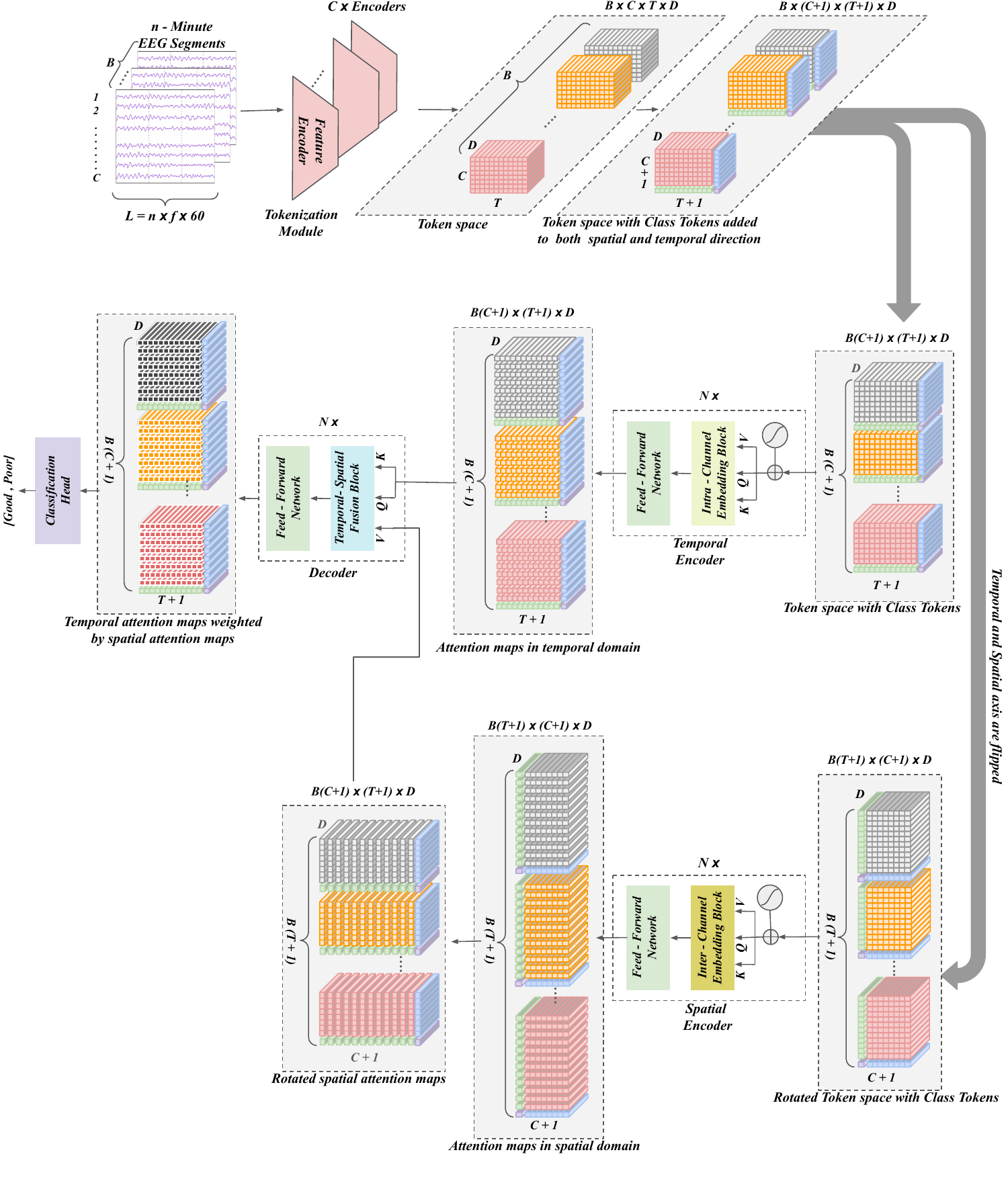}
\caption{\textit{Biaxialformer} framework: A processed $n$-minute segment is fed to \textit{Tokenization} block which converts each bipolar channel input to \textit{Tokens} preserving local patterns and spatial relationships. The information about the position of each token in the sequential feature space is encoded twice by the joint  \textit{Positional Encoding}. The \textit{Biaxialformer} focuses on relevant information between intra- and inter-channels by utilizing Temporal and Spatial encoders respectively. Outputs of the two encoders are then fused by the Decoder. Resulting attention maps are flattened and fed to the classification head for binary classification.}
\label{FIGURE1}
%\vspace{-1.0em}
\end{figure*}

As shown in Fig.\ref{FIGURE1}, the model consists of four main
components: (i) \textit{Tokenization Module} block encodes the high-frequency raw EEG signal information into token embeddings.
%\vspace{-1em}
(ii) \textit{Intra-channel embedding} block encodes the compressed and auxiliary augmented EEG data (with positional encoding and intra-channel class tokens) along the temporal axis. Subsequently, the intra-channel class tokens containing the temporal features within each EEG channel are generated. (iii) \textit{Inter-channel embedding} block encodes temporally segmented and auxiliary augmented EEG data (with positional encoding and inter-channel class tokens) along the channel axis.
Subsequently, inter-channel class tokens capture significant information about the relationships between the channels within each patch at a specified time. 
(iv) \textit{Temporal-spatial fusion }block effectively fuses intra- and inter-channel information to extract significant features from the EEG data.
%\vspace{-1em}
\subsection{Tokenization Module} \label{feature encoder}
\begin{figure*}[h]
\centering
\includegraphics[trim=0 170 0 0, clip, scale=0.65]{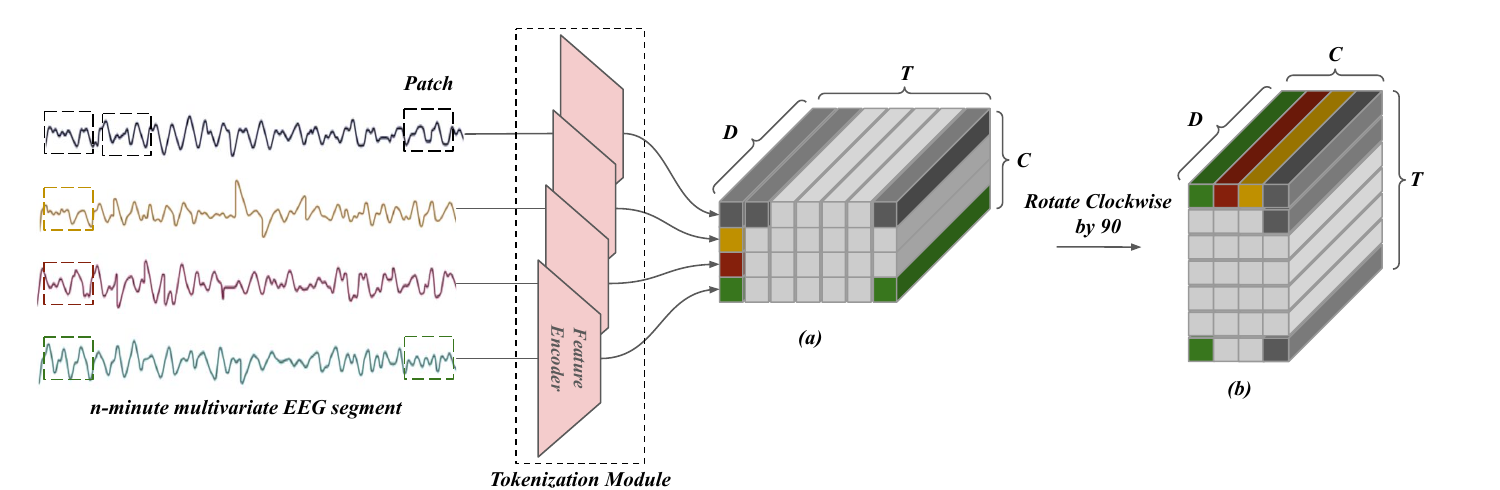}
\centering \caption{ Conversion of signal patches into token embeddings from a typical 4-channel EEG array as an example for two scenarios (a) intra-channel (temporal), and (b) inter-channel (spatial). Each EEG sequence (\textit{L}) in a \textit{n}-minute EEG segment is divided into patches. Based on stride (\textit{s}) and with the corresponding reception field (\textit{r}) of the feature encoder, each patch is converted into token embedding. In total \textit{T} tokens each with model dimension (\textit{D}) are generated from a \textit{n}-minute EEG segment.}
\label{TM}
%\vspace{-1em}
\end{figure*}

The input to the tokenization module is fed in terms of batches $B$ given as, 
\begin{equation}
\label{eqn2xb}
\mathbf{X}_{B}=\left\{\mathbf{\underset{1} {x}^{(t,k)}}, \mathbf{\underset{2} {x}^{(t,k)}}, \ldots, \mathbf{\underset{B} {x}^{(t,k)}} \right\}
\end{equation} where each input batch 
$\mathbf{X}_{B} \subseteq \mathbf{X}$ with batch size $ B\leq m $. The tokenization module then converts the signal patches into token embeddings by applying tokenization through a module architecture similar to the wav2vec\cite{baevski2020wav2vec}.
Our proposed architecture of tokenization comprises an independent feature encoder (FE) per channel. Hence we have $C$ feature encoders as shown in Fig. \ref{TM}.
%\vspace{-1em}
Each FE block is a stack of $l$ 1-D convolutional neural network (CNN) layers that excel at capturing local patterns and dependencies within the EEG channels w.r.t time, effectively reducing the data dimensionality while preserving relevant sequence-level semantic information. This feature extraction process transforms raw EEG signals into compact representations as tokens that the \textit{intra- and inter-channel embedding blocks} can further process. Tokenization allows the EEG data to be organized into discrete fragments, enabling the attention mechanism to model long-range dependencies and capture complex relationships within inter- and intra-EEG channels. The initial layer of FE includes an \textit{instance normalization } between the 1-D convolution and the \textit{Gaussian Error Linear Unit} (GELU) activation function, while the other layers
consist of a 1-D convolution followed by a GELU activation function. 

\begin{figure}[h]
\centering
\includegraphics[trim=0 410 310 20, clip, scale=0.65]{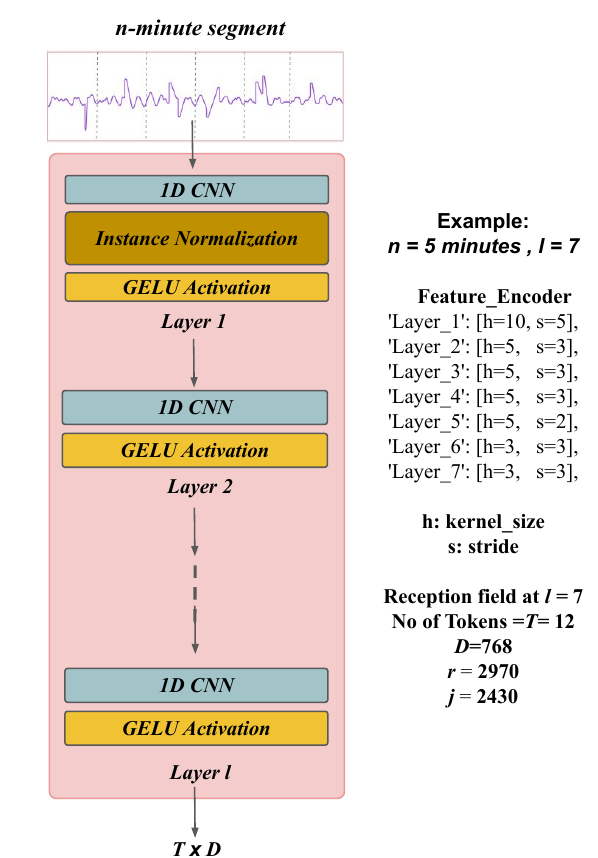}
\caption{Single FE design.}
\label{FE}
\end{figure}

\noindent The total context of the FE receptive field is a critical parameter in time-series analysis, as it determines the temporal context captured by each token in time-series modeling. In EEG signals, which are characterized by high temporal resolution and dynamic patterns, selecting an appropriate receptive field is essential to accurately capture relevant neural activity. A larger receptive field allows the model to capture long-range dependencies and broader contextual information, which is crucial for identifying patterns across extended periods. Conversely, a smaller receptive field focuses on fine-grained, localized patterns, which may be important for detecting transient or rapidly changing neural events. 
In this study, we design a varying receptive field by adjusting kernel sizes and strides to optimize the balance between capturing detailed, localized features and broader temporal dependencies, which is key to enhancing the model's ability to interpret and predict neural dynamics effectively. 
The total context of the FE receptive field ($r$) at the $l^{th}$ layer is given by (\ref{recpeqn}) \cite{araujo2019computing}
\begin{equation}
\label{recpeqn}
r_i=r_{i-1}+(h-1) \prod_{h=1}^{i-1} s_h  
\end{equation}
\noindent The first step to derive (\ref{recpeqn}) is to calculate the number of tokens (output feature map) for each layer. This is calculated by (\ref{tokeneqn}):
\begin{equation}
\label{tokeneqn}
 o_i=\frac{o_{i-1}+2 p-h}{s}+1
\end{equation}
where $ o_i$ is the number of the output features for layer $i$, $ o_{i-1}$ is the number of the input features, $p$ is the padding size, $h$ is the kernel size and $s$ is the stride of the layer $i$. Next, we need to calculate the jump ($j$) which in general, represents the cumulative stride. We can get $j$ by multiplying the strides of all layers before the current layer under investigation as follows:
\begin{equation}
\label{jumpeqn}
 j_i=j_{i-1}  * s 
\end{equation} where $ j_{i-1}$ is the jump of the previous layer. Finally, using previous values, we can calculate the size of the receptive field, using this formula:
 \begin{equation}
\label{recpeqn2}
r_i=r_{i-1}+(h-1) j_{i-1}
\end{equation}
The equation (\ref{recpeqn}) is the more general form of the formula (\ref{recpeqn2}) that is used for calculating the receptive field of $l^{th}$ layer.
It is to be noted that, each input univariate EEG series is converted into patches by the corresponding FE that can serve as input tokens to the attention mechanism. Given $s$ as stride between two consecutive patches and $D$  as token length (model dimension,768), each FE unfolds the $n$-minute EEG array  $\mathbf{X}_{B}$ into the embedded tensor $\mathbf{Z}_{B} \in \mathbb{R}^{B \times C \times T \times D}$ where $T$ is the number of patches, $T=\left\lfloor\frac{(L-D)}{s}\right\rfloor+1$.
%Also, the tokenization is performed using only 1-D CNNs in the temporal direction after reshaping EEG signals by aggregating the batch and channel axes. %The patches are mapped to the Transformer latent space of dimension D via trainable linear projection $\mathcal{W}_{\text {pos-inter }} \in \mathbb{R}^{B(T+1) \times (C+1) \times D}$,
%So, after performing the tokenization ($f: \mathbb{R}^{BC \times N \times P} \rightarrow  \mathbb{R}^{BC \times T \times D}$, $D$ indicates the number of embedded dimensions and $T$  is the length of the token), we split the data on both the batch and channel axes, which were combined. Therefore, we could generate the embedded dimensions ($V=f(x) \in \mathbb{R}^{B \times C \times T \times D}$)  from the EEG time-series.
For instance, the proposed tokenization module comprised seven embedding layers with the following kernel sizes of [10, 5, 5, 5, 5, 3, 3] and strides of [5, 3, 3, 3, 2, 3, 3] for $r$ corresponding to $\sim$30 seconds at the 100 Hz sample rate. Here values of $r$ and $j$ were 2970 and 2430 in samples respectively. Hence 12 tokens were generated from each channel in a 5-minute segment EEG array as shown in Fig. \ref{FE}.
%\vspace{-1em}
%\vspace{-2.em}
\subsection{Positional encoding}
The attention mechanism alone can not explicitly model the sequence token order (i.e., time steps or channels). Therefore, we employ a learnable additive positional encoding scheme to provide positional information about the tokens. Two separate positional encodings and class tokens in the intra- and inter-channel axes are added to encode temporal and spatial information respectively. The joint learning occurs as both sets of positional encodings (temporal and spatial) are learned together as part of the overall model’s training process. The proposed joint learning of positional encodings addresses the challenge of \textit{inter-channel correlation forgetting} \cite{han2024mcformer} in MTS analysis, such as EEG signals, by ensuring that spatial dependencies between channels are explicitly modeled alongside temporal dynamics. Traditional MTS models often focus on temporal relationships within each sequence (or channel), leading to a neglect of spatial correlations between channels, especially when they are processed independently. This can result in channel correlation forgetting, where important inter-channel interactions, such as those critical in EEG data (e.g., neural oscillations or network activities spanning multiple electrodes), are lost. At the same time, this architecture promotes effective \textit{CI} by ensuring that each channel’s temporal sequence is modeled independently via intra-channel encoding, allowing the model to recognize unique patterns within each channel. Since EEG arrays are MTS representations, we assumed that information on the temporal order is significant for comprehending the relationship between channels. 
So, we first append $B \times C$ number of class tokens (intra-channel class tokens), $\mathrm{CLS}_{\text {intra }} \in \mathbb{R}^{B \times C \times D}$, at the beginning of the temporal axis, followed by $B \times (T + 1)$ number of class tokens (inter-channel class tokens), $\mathrm{CLS}_{\text {inter }} \in \mathbb{R}^{B \times (T+1) \times D}$ along channel axis. Once both class tokens are appended, we then add intra-channel-wise learnable positional encoding, $\mathcal{W}_{\text {pos-intra }} \in \mathbb{R}^{B\times C \times (T+1) \times D}$, at the beginning of the temporal axis. 
This temporally divided data, with
the dimension $B\times C \times (T + 1) \times D$ is then subjected to inter-channel-wise learnable  positional encoding $\mathcal{W}_{\text {pos-inter }} \in \mathbb{R}^{B \times (C+1) \times (T+1) \times D}$, along the channel axis.  The overall shape of the resultant feature maps is $B \times (C + 1) \times (T + 1) \times  D$. The positional encoding and class tokens are added simultaneously to ensure that, the data size remains consistent before and after processing. The resultant feature maps will be further subjected to two-stage multi-head attention (MHA) separately.
\vspace{-1.35em}
\subsection{\textit{Intra Channel Embedding Block (Temporal)}}This embedding block utilizes a native Transformer encoder design \cite{dosovitskiy2020image,vaswani2017attention} to model the EEG signals' temporal features. This block employs MHA and a feed-forward network on channel-wise-divided data to encode significant intra-channel information. To achieve this the channel and batch axis of the feature maps were
merged resulting in dimension $B(C + 1) \times (T + 1) \times  D$. This results in a reshaped representation where each EEG channel is treated independently with its corresponding temporal data for further processing.
The MHA mechanism focuses on learning temporal dependencies within each channel to build patch-wise attention. For this, the data is projected into three distinct subspaces to compute attention scores using learnable matrices $Q_{T}, K_{T},$ and $V_{T}$, where each corresponds to the temporal patch tokens (time-steps). \textit{Query} $(Q_{T})$: a projection of the input data that defines the information the model seeks in the temporal dimension. \textit{Key} $(K_{T})$: a second projection that represents the potential locations where relevant information might be found, i.e., temporal positions in the sequence. \textit{Value}  $(V_{T})$: a third projection containing the actual content or features at each time step (patch) that are to be retrieved based on the query-key matching process.
These matrices are learned during training and can be described as:
\vspace{-0.6em}
\begin{equation}
  Q_{T} = Z_{T}\mathcal{W}_{Q_{T}}, K_{T} = Z_{T}\mathcal{W}_{K_{T}}, V_{T} = Z_{T}\mathcal{W}_{V_{T}}   
\end{equation} where $Z_{T}$ is the input temporal data and $\mathcal{W}_{Q_{T}},\mathcal{W}_{K_{T}}, \mathcal{W}_{V_{T}}$
are the learned weight matrices that project the input into the query, key, and value spaces, respectively.

By aggregating these dependencies into class tokens, the model efficiently encodes the temporal information within each channel, ready for further processing and integration with spatial features from other channels.
\vspace{-1em}
\subsection{\textit{Inter Channel Embedding Block (Spatial)}} 
This embedding block also utilizes a native Transformer encoder design to model the EEG signals' spatial features.
This block focuses on capturing spatial relationships between EEG channels, i.e., inter-channel dependencies within the same temporal slice or patch. After the intra-channel (temporal ) encoding is completed, the original feature maps are restructured by merging the temporal and batch axes, resulting in a transformation with dimensions $B(T + 1) \times (C + 1) \times  D$. This transformation enables the model to treat each time step (or patch) across all channels as a single entity and apply attention to learn the inter-channel dependencies. Now, the inter-channel-wise MHA mechanism and a feed-forward network are employed on this resultant temporally divided data, to extract significant information between the EEG channels within a specific time. Similar to the intra-channel attention, three learnable matrices — $Q_{S}, K_{S},$ and $V_{S}$ are used to construct attention scores, but this time across channels at a specific temporal slice, rather than across time steps within a channel. The learnable matrices are defined as: \textit{Query} $(Q_{S})$: represents the current EEG channel, indicating which inter-channel relationships to focus on within the current temporal slice. \textit{Key} $(K_{S})$: represents the potential spatial relationships between channels, i.e., how one EEG channel’s information can be related to others. \textit{Value}  $(V_{S})$: holds the actual content or features of each channel, which will be weighted based on how relevant each channel is to the others. The matrices are calculated as:
\vspace{-0.6em}
\begin{equation}
   Q_{S} = Z_{S}\mathcal{W}_{Q_{S}}, K_{S} = Z_{S}\mathcal{W}_{K_{S}}, V_{S} = Z_{S}\mathcal{W}_{V_{S}}  
\end{equation} where $Z_{S}$  represents the transformed feature maps where each token corresponds to an EEG channel at a specific time step,  and $\mathcal{W}_{Q_{S}},\mathcal{W}_{K_{S}}, \mathcal{W}_{V_{S}}$
are the learned weight matrices. After computing the inter-channel-wise attention, the model aggregates the spatial dependencies between EEG channels into inter-channel class tokens. Each class token represents a summary of the relationships between the channels for a specific temporal slice, encapsulating the spatial structure of the EEG data at that time.

Thus, by employing and operating two-stage attention mechanisms jointly, we could encode inter- and intra-channel information from the downsampled EEG patches. The two embedding blocks operate independently using two different types of class tokens. The reuse of feature maps derived from tokenization undergoes a reshape operation to seamlessly transition the two-stage MHA operations for inter- and intra-channel information respectively. By this, we achieve devoid of direct
inter-informational exchange, potentially resulting in a lack of coordinated feature extraction across
channels and sequences. This intricate design aims to establish a cohesive interplay between temporal and spatial representations, ensuring a harmonized extraction of EEG features.
\vspace{-1.em}
\subsection{\textit{Temporal-Spatial Fusion Block (Decoder)}} This block is designed to integrate the outputs from the intra-channel (temporal) and inter-channel (spatial) encoders through a \textit{Cross-attention} ($CR$) mechanism. This $CR$ process leverages the information encoded separately by the temporal and spatial encoders to create a unified representation of the EEG data, combining both time and channel relationships. The model then effectively integrates temporal and spatial information through MHA, producing a fused representation that captures the critical relationships between EEG channels and their temporal evolution.
The decoder's $CR$ operates by mapping the Values from the spatial encoder and the Query-Key pairs from the temporal encoder to compute the final fused output. The attention mechanism is built using three learnable matrices:
\textit{Query} $(Q_{ST})$: extracted from the temporal encoder’s output, representing the temporal features (intra-channel information). \textit{Key} $(K_{ST})$: also derived from the temporal encoder, providing the reference points that determine which temporal features are relevant to the spatial information. \textit{Value}  $(V_{ST})$: taken from the spatial encoder’s output, representing the inter-channel (spatial) information to be retrieved based on the temporal query-key matching. These matrices are defined as:
\begin{equation}
    Q_{ST} = Z_{to}\mathcal{W}_{Q_{ST}}, K_{ST} = Z_{to}\mathcal{W}_{K_{ST}}, V_{ST} = Z_{so}\mathcal{W}_{V_{ST}} 
\end{equation} where $Z_{to}$  is the output feature maps from the FE and,  and $Z_{so}$  is output from the spatial encoder. $\mathcal{W}_{Q_{ST}},\mathcal{W}_{K_{ST}}, \mathcal{W}_{V_{ST}}$
are the learned weight matrices.\\
The final output is computed as a weighted sum of the values, where the weight assigned to each value is computed by a compatibility function of the query with the corresponding key. The compatibility function computes attention scores by taking the dot product of the Query (from the temporal encoder) and Key (also from the temporal encoder), as follows:
%\vspace{-1em}
\begin{equation}
\label{eqn11}
\operatorname{CR}\left(Q_{ST}, K_{ST}, V_{ST}\right)=\operatorname{Softmax}\left(\frac{Q_{ST}\left(K_{ST}\right)^T}{\sqrt{d_k}}\right) V_{ST}
\end{equation}
Here, $\sqrt{d_k}$ is the dimension of the Key, ensuring the dot product is scaled appropriately for numerical stability. The attention scores represent how well each temporal Query aligns with the Keys, determining which temporal features are relevant to the spatial features encoded in the Value matrix. Once the attention scores are calculated, they are applied to the Value matrix (from the spatial encoder) to compute a weighted sum. This weighting allows the model to focus on the spatial features that are most relevant to the temporal features, effectively fusing the temporal and spatial information.
\vspace{-1em}
\subsection{Classification head}
The classification head converts high-dimensional decoder representations into a final prognosis of either 'poor' or 'good.' It receives weighted attention maps that capture temporal and spatial relationships within the EEG data. These maps are flattened into a one-dimensional vector of size $(C+1) \times (T+1) \times D$, then passed through a fully connected layer. A softmax layer applied to the logits produces probabilities reflecting the model’s confidence in each prognosis category.

\section{Experimental Setup}

\subsection{Scoring Metric}
The scoring metric (SM) is TPR for predicting a poor outcome (CPC of 3, 4, or 5) given an FPR $\leq$ 0.05 at 72 hours after the ROSC. 
\begin{equation}
   {SM} = {\frac{{TP}}{{TP} + {FN}} \Big|}_{FPR \leq 0.05}  , \textit{72h after ROSC} %\frac{\text{FP}}{\text{FP} + \text{TN}} < 0.05 
\end{equation} The clinical rationale for evaluating the algorithm’s performance using the TPR at a FPR of 0.05 is as follows. In clinical settings, prognostic assessments play a crucial role in deciding whether to continue life-supporting interventions. A false-positive prediction of poor outcome is particularly concerning, as it may lead to the withdrawal of life support from a patient who has the potential to recover consciousness, resulting in their death. While false negatives—failing to predict a poor outcome, thereby prolonging life support in patients who ultimately have a poor prognosis—are also challenging, they are generally considered less critical than false positives. Consequently, professional society guidelines recommend that prognostic tests maintain a FPR of 5\% or lower.

There are two key reasons for assessing the algorithm’s performance at 72 hours, rather than at earlier time points. First, research suggests that EEG trends over time provide valuable prognostic information \cite{zheng2021predicting1}. Second, premature predictions of poor neurologic outcomes may lead to self-fulfilling prophecies. Therefore, it is recommended to delay formal neurologic prognoses until at least 72 hours post-event.
\vspace{-1em}
\subsection{Implementation details \& Evaluation Settings}
 Adam optimizer is applied to update the model's weights with a batch size of 10 and a learning rate 0.0001 for 40,000 iterations. A cosine annealing schedule is employed to decay the learning rate during training. For the binary classification task of the neuro-prognostication, we utilized the binary cross-entropy loss function to guide the training, tailored to the 'Good' and 'Poor' neurological outcome labels: \vspace{-0.5em}\begin{equation}
L_{(\hat{y}, y)} = -\frac{1}{N}\sum_{i=1}^{N} \left[ y_i \log(\hat{y}_i) + (1-y_i) \log(1-\hat{y}_i) \right]
\end{equation} where \(N\) is the batch size, \(y_i\) is the true label (\(y_i = 1\) for 'Good' and \(y_i = 0\) for 'Poor'), and \(\hat{y}_i\) is the predicted outcome for the \(i\)-th patient.

The inherent attention mechanism of the Transformer-based implementation typically requires a significant amount of training data to capture intrinsic correlations among EEGs. This is achieved by deploying a segmentation approach for the selected hour of EEG recording into 5-minute segments. To further robust the training process, each iteration is started by randomly selecting an hour of EEG recording of a patient, followed by a segmentation strategy, and then selecting a random 5-minute segment from the corresponding chosen random hour. Using this random selection and 5-minute segmentation techniques, we could generate $\sim$0.5 million training samples from EEG recordings of 607 patients.

\section{Results \& Discussion}
\subsection{Classification Results}

The proposed study used the publicly shared training dataset of 607 patients from five hospital centers (A, B, D, E, F) for the experiments in a leave-one-out center format. Notably, hospital A accounts for half of the patients in the dataset. Since the original test dataset was not publicly made available after the challenge period, this study initiated a comprehensive experimental regimen, with the formulation of test-train splits. A cross-validation was performed individually on each of the five hospitals within the public training dataset. Consequently, each iteration of the in-house test set comprised solely one hospital, which had been held out from the training set. Subsequently, we compiled and reported the cross-validation results obtained for all participating hospitals across different time windows (in hours 12, 24, 48, 72), including the average results of various other performance metrics in Tables \ref{tab:windowresults}, \ref{tab:PMresults}.
%\vspace{-1em}
\begin{table}[h]
\caption{\label{tab:windowresults} Assessment of Biaxialformer using SM (TPR at FPR of 0.05) in leave-one-center-out format for different time-windows}
\centering
\begin{tabular}{|l|l|l|l|l|}
\hline
\rowcolor{lightgray} \begin{tabular}[c]{@{}l@{}}Hospital\\ \end{tabular} & \begin{tabular}[c]{@{}l@{}}SM 12h\end{tabular} & \begin{tabular}[c]{@{}l@{}}SM 24h\end{tabular} & \begin{tabular}[c]{@{}l@{}}SM 48h\end{tabular} & \begin{tabular}[c]{@{}l@{}}SM 72h\end{tabular} \\ \hline
A                                                       & \multicolumn{1}{c|}{0.1742}                                                                                                                &    \multicolumn{1}{c|}{0.3810}                                                                                                             &        \multicolumn{1}{c|}{0.5120}                                                                                                         &         \multicolumn{1}{c|}{0.4820}                                                       \\ \hline
B                                                       &     \multicolumn{1}{c|}{0.1650}                                                                                                            &        \multicolumn{1}{c|}{0.2466}                                                         &        \multicolumn{1}{c|}{0.3270}                                                         &     \multicolumn{1}{c|}{0.2440}                                                            \\ \hline
D                                                       & \multicolumn{1}{c|}{0.1856}                                     & \multicolumn{1}{c|}{0.3962}                                    & \multicolumn{1}{c|}{0.5792}                                    & \multicolumn{1}{c|}{0.5144}                                    \\ \hline
\textbf{E}                                            & \multicolumn{1}{c|}{0.1860}                                                                 & \multicolumn{1}{c|}{0.3662}                                                               &  \multicolumn{1}{c|}{0.8204}                                                         &  \multicolumn{1}{c|} {\textbf{0.7798}}                                                               \\ \hline
F                                                     & \multicolumn{1}{c|}{0.1200}                                     & \multicolumn{1}{c|}{0.4846}                                    & \multicolumn{1}{c|}{0.5802}                                    & \multicolumn{1}{c|}{0.5619}                       \\ \hline
%Avg.                                                    & \multicolumn{1}{c|}{0.1200}                                     & \multicolumn{1}{c|}{0.4846}                                    & \multicolumn{1}{c|}{0.5102}                                    & \multicolumn{1}{c|}{0.5619}                       \\ \hline
\end{tabular}
\end{table} 
\vspace{-1em}
It is to be noted that throughout our study we chose the EEG segment duration to be 5 minutes, with $r$= 30 seconds. As shown in Table \ref{tab:windowresults}, the SM is higher for the time window of 48 hours across all hospitals. This is attributed to the negatively skewed data distribution in the study ( refer Fig.2 of \cite{amorim2023international} ). Specifically, most patients across the five hospitals have data recorded for at least a day but less than three days.
%\vspace{-1em}
%\vspace{-1em}
Further, Tables \ref{tab:windowresults}, \ref{tab:PMresults} show that hospitals D, E, and F performed well and consistently produced the expected results. However, hospital A had a slightly lower SM of 0.4820. Since hospital A represents about 43\% of the study population, the model was consequently trained on slightly fewer patients compared to the other four hospitals, which likely contributed to its reduced performance in this case. %Hospital B's performance also deviated from the others, possibly due to unique patient features in its data. 
While Hospital B’s SM may seem low, its other performance metrics still reveal good predictive performance. The lower custom SM, focused on a high TPR at an FPR $\leq$ 5\%, emphasizes minimizing false alarms, and doesn’t fully reflect the model’s ability to capture positive cases effectively. Given the possibility of unique patient characteristics in Hospital B's data, these alternative metrics underscore its reliability despite the stricter scoring criterion.
%\vspace{-1.25em}
\begin{table}[h]
\caption{\label{tab:PMresults} Assessment of Biaxialformer using other performance metrics in leave-one-center-out format}
\centering
\begin{tabular}{|l|l|l|l|l|}
\hline
\rowcolor{lightgray} \begin{tabular}[c]{@{}l@{}}Hospital\\ \end{tabular} & \begin{tabular}[c]{@{}l@{}}SM 72h\end{tabular} & \begin{tabular}[c]{@{}l@{}}F1 \\ \end{tabular} & \begin{tabular}[c]{@{}l@{}}AUROC\\ \end{tabular} & \begin{tabular}[c]{@{}l@{}}AUPRC\\ \end{tabular} \\ \hline
A                                                       &    \multicolumn{1}{c|}{0.4820}                                                                                 &     \multicolumn{1}{c|}{0.6590}                                                                                &     \multicolumn{1}{c|}{0.7980}                                                                                &            \multicolumn{1}{c|}{0.8930}                                                                         \\ \hline
B                                                       &  \multicolumn{1}{c|}{{0.2440}}                                                               &            \multicolumn{1}{c|}{0.5200}                                                     &                   \multicolumn{1}{c|}{0.6210}                                              &       \multicolumn{1}{c|}{0.7930}                                                          \\ \hline
D                                                       & \multicolumn{1}{c|}{0.5144}                                     & \multicolumn{1}{c|}{0.6877}                                    & \multicolumn{1}{c|}{0.8210}                                    & \multicolumn{1}{c|}{0.9010}                                    \\ \hline
E                                                       &  \multicolumn{1}{c|}{\textbf{0.7798}}                                                              & \multicolumn{1}{c|}{0.7660}                                                                &    \multicolumn{1}{c|}{0.8490}                                                            &    \multicolumn{1}{c|}{0.9440}                                                            \\ \hline
F                                                      & \multicolumn{1}{c|}{0.5619}                                     & \multicolumn{1}{c|}{0.6912}                                    & \multicolumn{1}{c|}{0.7619}                                    & \multicolumn{1}{c|}{0.8012}                            \\ \hline
Avg.                                                     & \multicolumn{1}{c|}{\textbf{0.5164}}                                     & \multicolumn{1}{c|}{0.6518}                                    & \multicolumn{1}{c|}{0.7688}                                    & \multicolumn{1}{c|}{0.8643}                       \\ \hline
\end{tabular}
\end{table} 
\vspace{-2.85em}
\subsection{Ablations}
To demonstrate and validate the performance of Biaxialformer, we conducted targeted ablation studies, maintaining consistent model hyperparameters across all experiments. Hospital \textit{E} was selected for these ablations and baseline comparisons, as it yielded the most favorable results in our cross-validation setup.
%To highlight and justify the performance of the proposed Biaxialformer, we  performed subsequent ablations. Model hyperparameters were kept consistent throughout the experiments. It is to be noted that we chose hospital \textit{E} for all our ablation experiments and baseline studies since we obtained the best results for the \textit{E}  in our cross-validation setup. 
\subsubsection{Two-stage MHA architecture} 
To verify the performance of the proposed two-stage MHA architecture in \textit{Biaxialformer}, we conducted ablation studies that analyzed the individual contributions of the temporal and spatial encoder blocks. These experiments involved removing components of the two-stage MHA  and evaluating their performance separately. 
%\vspace{-1.25em}
\begin{table}[h]
\caption{\label{tab:intraresults} Intra-channel (Temporal) block results on Hospital E}
\centering
\begin{tabular}{|l|l|l|l|l|}
\hline
\rowcolor{lightgray} \begin{tabular}[c]{@{}l@{}} Window\end{tabular} & \begin{tabular}[c]{@{}l@{}}SM\\ \end{tabular} & \begin{tabular}[c]{@{}l@{}}F1 \\ \end{tabular} & \begin{tabular}[c]{@{}l@{}}AUROC \\ \end{tabular} & \begin{tabular}[c]{@{}l@{}}AUPRC \\ \end{tabular} \\ \hline
12h                                                       & \multicolumn{1}{c|}{0.1642}                                                                                                                &    \multicolumn{1}{c|}{0.4810}                                                                                                             &        \multicolumn{1}{c|}{0.7120}                                                                                                         &         \multicolumn{1}{c|}{0.7820}                                                       \\ \hline
24h                                                      &     \multicolumn{1}{c|}{0.3165}                                                                                                            &        \multicolumn{1}{c|}{0.5466}                                                         &        \multicolumn{1}{c|}{0.8070}                                                         &     \multicolumn{1}{c|}{0.8440}                                                            \\ \hline
48h                                                       & \multicolumn{1}{c|}{0.6856}                                     & \multicolumn{1}{c|}{0.6962}                                    & \multicolumn{1}{c|}{0.8792}                                    & \multicolumn{1}{c|}{0.9144}                                    \\ \hline
\textbf{72h}                                            & \multicolumn{1}{c|}{0.5860}                                                                 & \multicolumn{1}{c|}{0.7162}                                                               &  \multicolumn{1}{c|}{0.8433}                                                         &  \multicolumn{1}{c|}{0.8710}                                                               \\ \hline
%F                                                     & \multicolumn{1}{c|}{0.1200}                                     & \multicolumn{1}{c|}{0.4846}                                    & \multicolumn{1}{c|}{0.5102}                                    & \multicolumn{1}{c|}{0.5619}                       \\ \hline
%Avg.                                                    & \multicolumn{1}{c|}{0.1200}                                     & \multicolumn{1}{c|}{0.4846}                                    & \multicolumn{1}{c|}{0.5102}                                    & \multicolumn{1}{c|}{0.5619}                       \\ \hline
\end{tabular}
\end{table}
%\vspace{-1em}
\begin{table}[h]
\caption{\label{tab:interresults} Inter-channel (Spatial) block results on Hospital E}
\centering
\begin{tabular}{|l|l|l|l|l|}
\hline
\rowcolor{lightgray} \begin{tabular}[c]{@{}l@{}} Window\end{tabular} & \begin{tabular}[c]{@{}l@{}}SM\\ \end{tabular} & \begin{tabular}[c]{@{}l@{}}F1\\ \end{tabular} & \begin{tabular}[c]{@{}l@{}}AUROC \\ \end{tabular} & \begin{tabular}[c]{@{}l@{}}AUPRC \\ \end{tabular} \\ \hline
12h                                                       & \multicolumn{1}{c|}{0.0932}                                                                                                                &    \multicolumn{1}{c|}{0.4760}                                                                                                             &        \multicolumn{1}{c|}{0.5937}                                                                                                         &         \multicolumn{1}{c|}{0.6120}                                                       \\ \hline
24h                                                      &     \multicolumn{1}{c|}{0.2995}                                                                                                            &        \multicolumn{1}{c|}{0.5376}                                                         &        \multicolumn{1}{c|}{0.7470}                                                         &     \multicolumn{1}{c|}{0.7961}                                                            \\ \hline
48h                                                       & \multicolumn{1}{c|}{0.5849}                                     & \multicolumn{1}{c|}{0.6892}                                    & \multicolumn{1}{c|}{0.8672}                                    & \multicolumn{1}{c|}{0.8944}                                    \\ \hline
\textbf{72h}                                            & \multicolumn{1}{c|}{0.4761}                                                                 & \multicolumn{1}{c|}{0.5962}                                                               &  \multicolumn{1}{c|}{0.8173}                                                         &  \multicolumn{1}{c|}{0.8413}                                                               \\ \hline
\end{tabular}
\end{table}
\vspace{-1em}
Specifically, we trained two separate transformer encoder models: one using only the intra-channel embedding block (temporal encoder) and the other using only the inter-channel embedding block (spatial encoder). In the first ablation, the \textit{Biaxialformer} was trained using only the intra-channel block, excluding the inter-channel encoder. In the second setup, only the inter-channel block was considered. The results of these studies, as shown in Tables \ref{tab:intraresults} and \ref{tab:interresults} for Hospital \textit{E}, demonstrate the individual contributions of temporal and spatial features, respectively. The results indicated that the full two-stage MHA architecture, with the temporal-spatial fusion block, significantly outperformed both individual ablations.
Furthermore, the intra-channel embedding block outperformed the inter-channel embedding block, highlighting 
the difference in the amount of tokenization between them. In the intra-channel embedding block, the model tokenizes each time series (channel) into smaller temporal segments, and this fine-grained tokenization results in a larger number of tokens per channel, allowing the model to capture more detailed temporal patterns, including short-term dependencies like burst suppression, oscillatory patterns, and transient events.

The inter-channel embedding block, in contrast, focuses on modeling spatial relationships between different EEG channels. Here, the tokenization is typically coarser because the number of EEG channels is far fewer as compared to the length of time series data within each channel.
While spatial dependencies in EEG (e.g., coherence between electrodes) are important, they don't require as many tokens because spatial interactions tend to evolve more slowly and are often more stable than temporal dynamics. Therefore, the inter-channel tokenization yields fewer tokens than the intra-channel block, limiting its overall contribution to the model's performance.
%\vspace{-1em}
\subsubsection{Conditioning of Decoder}
In the original \textit{Biaxialformer} architecture, the \textit{Value}  $(V_{ST})$ of the decoder is conditioned using $Z_{so}$. In this ablation, we condition the decoder on $Z_{to}$ instead. Specifically, the cross-attention in Eq. (\ref{eqn11}) for the decoder is altered by replacing the $V_{ST}$ input with $Z_{to}$, ensuring that the weighting of the attention maps for the final result is now focused on the temporal dimension.
%\vspace{-1.em}
\begin{table}[h]
\caption{\label{tab:weightresults} Weighting from temporal and spatial encoders on hospital E}
\centering
\begin{tabular}{|l|l|l|l|l|}
\hline
\rowcolor{lightgray} \begin{tabular}[c]{@{}l@{}} Conditioning \end{tabular} & \begin{tabular}[c]{@{}l@{}}SM (72h) \end{tabular} & \begin{tabular}[c]{@{}l@{}}F1 \\ \end{tabular} & \begin{tabular}[c]{@{}l@{}}AUROC \\ \end{tabular} & \begin{tabular}[c]{@{}l@{}}AUPRC \\ \end{tabular} \\ \hline

\begin{tabular}[c]{@{}l@{}}Spatial weighted 
 \\ by Temporal \end{tabular}                                                      & \multicolumn{1}{c|}{0.5762}                                     & \multicolumn{1}{c|}{0.6994}                                    & \multicolumn{1}{c|}{0.7972}                                    & \multicolumn{1}{c|}{0.8746}                                    \\ \hline

\begin{tabular}[c]{@{}l@{}}Temporal weighted 
 \\ by Spatial \end{tabular}                                                   & \multicolumn{1}{c|}{0.7798}                                     & \multicolumn{1}{c|}{0.7660}                                    & \multicolumn{1}{c|}{0.8490}                                    & \multicolumn{1}{c|}{0.9440}                                    \\ \hline
\end{tabular}
\end{table}
\vspace{-1em}
Table \ref{tab:weightresults} furnish comparative results of this ablation versus original  \textit{Biaxialformer}. Our findings indicate that conditioning the decoder on $Z_{so}$ yielded better results than $Z_{to}$. This can be attributed to the importance of spatial encoding in EEG data, where interactions between different brain regions (captured by spatial encoding) are crucial for tasks like neurological outcome prediction. Focusing on spatial features, the model can effectively capture inter-channel dependencies, leading to improved attention weighting and better performance. In contrast, relying solely on temporal features may overlook these critical spatial dynamics, reducing prediction accuracy.
%\vspace{-1em}
\subsubsection{Segment Variability \& Window Size} 
We conducted ablation studies exploring both window size and segment variability to assess the impact of segment selection and its duration on model performance. Initially, we tested the effect of varying the choice of window size from a given hour to 3, 10, and 12 minutes, instead of the 5 minutes used throughout the study.
Table \ref{tab:segments} shows the optimal 5-minute window in the Biaxialformer architecture likely balances capturing temporal context and computational efficiency. While longer lookback windows, such as 10 or 12 minutes, can identify broader dependencies, they may dilute critical short-term features, increase computational costs, and risk overfitting or noise. Conversely, a 3-minute window may miss critical context, limiting feature extraction. We expanded this analysis by testing the effect of using multiple 5-minute segments from various hours per patient and aggregating predictions using a \textit{mode} operation, revealing how varying the number of hours influences the SM for Hospital E. 
%\vspace{-1.35em}
\begin{table}[h]
\caption{\label{tab:segments} Segment Variability \& Window Size on hospital E}

\begin{tabular}{|c|c|c|c|c|c|}
\hline
\begin{tabular}[c]{@{}l@{}} Window Size \\ (in mins) \end{tabular} & \begin{tabular}[c]{@{}l@{}}-\end{tabular} & \begin{tabular}[c]{@{}l@{}}3\end{tabular} & \begin{tabular}[c]{@{}l@{}}5 \\ \end{tabular} & \begin{tabular}[c]{@{}l@{}}10\\ \end{tabular} & \begin{tabular}[c]{@{}l@{}}12\\ \end{tabular} \\ \hline

\begin{tabular}[c]{@{}l@{}}SM (72h) \end{tabular}          & \multicolumn{1}{c|}{-}                                             & \multicolumn{1}{c|}{0.7152}                                     & \multicolumn{1}{c|}{0.7798}                                    & \multicolumn{1}{c|}{0.2319}                                    & \multicolumn{1}{c|}{0.1990}                                    \\ \hline
\begin{tabular}[c]{@{}l@{}} No.Segments \\ (5 min) \end{tabular} & \begin{tabular}[c]{@{}l@{}}1 \end{tabular} & \begin{tabular}[c]{@{}l@{}}5 \end{tabular} & \begin{tabular}[c]{@{}l@{}}7 \\ \end{tabular} & \begin{tabular}[c]{@{}l@{}}9\\ \end{tabular} & \begin{tabular}[c]{@{}l@{}}11\\ \end{tabular} \\ \hline

\begin{tabular}[c]{@{}l@{}}SM (72h) \end{tabular}          & \multicolumn{1}{c|}{0.7798}                                             & \multicolumn{1}{c|}{0.8131}                                     & \multicolumn{1}{c|}{0.8255}                                    & \multicolumn{1}{c|}{0.8119}                                    & \multicolumn{1}{c|}{0.8012}                                   \\ \hline
\end{tabular}
\end{table} 
%\vspace{-1em} 
Table \ref{tab:segments} show that increasing the segment count improves the SM by capturing more representative temporal patterns. However, beyond a certain point, SM tapers off, indicating an optimal balance between EEG variability and computational efficiency, highlighting that strategic segment selection enhances robustness without redundancy. 
%\vspace{-1.35em}
\begin{table}[h]
\centering
\caption{\label{tab:RFE} Exploration of reception field on hospital E}
\begin{tabular}{|c|c|c|c|c|c|}
\hline
\rowcolor{lightgray} \multicolumn{1}{|l|}{\begin{tabular}[c]{@{}l@{}}Duration \\ of \\ $r$(secs)\end{tabular}} & \multicolumn{1}{l|}{\begin{tabular}[c]{@{}l@{}}30 \\ secs\\ \\  $r$=2970\\$j$=2430\\$o$ =12 \end{tabular}} & \multicolumn{1}{l|}{\begin{tabular}[c]{@{}l@{}}20 \\ secs\\ \\ $r$=2160 \\ $j$ =2750\\ $o$ =13\end{tabular}} & \multicolumn{1}{l|}{\begin{tabular}[c]{@{}l@{}}15\\ secs\\ \\ $r$=1620\\ $j$=2070\\$o$ =18 \end{tabular}} & \multicolumn{1}{l|}{\begin{tabular}[c]{@{}l@{}}10\\ secs\\ \\ $r$=1080\\ $j$=1190\\$o$ =27\end{tabular}} & \multicolumn{1}{l|}{\begin{tabular}[c]{@{}l@{}}5\\ secs\\ \\ $r$=480\\ $j$=1050\\$o$ =61\end{tabular}} \\ \hline
\multirow{7}{*}{\begin{tabular}[c]{@{}c@{}}Tokeni-\\zation \\ Module \\ $(h,s)$\end{tabular}}& 10,5 & 10,5 & 10,5 & 10,5  & 10,5  \\ \cline{2-6} 
& 5,3 & 5,4   & 5,3   & 5,2   & 5,3 \\ \cline{2-6} 
& 5,3 & 5,3    & 5,3   & 5,2   & 5,2 \\ \cline{2-6} 
& 5,3 & 5,2    & 5,2   & 5,2    & 5,2 \\ \cline{2-6} 
& 5,2 & 5,2    & 5,2   & 3,3     & 3,2  \\ \cline{2-6} 
& 3,3  & 3,3    & 3,3   & 3,3     & 3,2  \\ \cline{2-6} 
& 3,3 & 3,3    & 3,3   & 3,3      & 3,2  \\ \hline
\multicolumn{1}{|l|}{\begin{tabular}[c]{@{}l@{}}SM 72h\end{tabular}}  & \multicolumn{1}{l|}{0.7798}                   & \multicolumn{1}{l|}{0.7912}                                                                  & \multicolumn{1}{l|}{0.8471}                                                                  & \multicolumn{1}{l|}{0.7664}                                                                  & \multicolumn{1}{l|}{0.6823}                                                                \\ \hline
\end{tabular}
\end{table}
%\vspace{-1em}
\subsubsection{Exploration of reception field ($r$)}
We systematically explored the receptive field $r$ of FE by varying kernel sizes and strides, resulting in different tokenization schemes. Each scheme produced distinct tokens, with $r$ defined in terms of sample durations (seconds) for EEG signals. This ablation experiment aimed to identify the optimal $r$ value that maximizes model performance by balancing the granularity of tokenization with the temporal context captured in each token. Table \ref{tab:RFE} highlights classification results from Hospital $E$, with five different $r$ values tested using a fixed window size of 5 minutes. Notably, an optimal value of $r$ = 15 seconds, corresponding to 1500 samples ($f_s = 100 Hz$), produced the best results, striking a balance between detail and broader context. With smaller 
$r$ values (5 or 10 seconds), token counts increased (61 and 27 tokens, respectively), potentially leading to over-segmentation and loss of contextual information, despite retaining high granularity. Conversely, larger 
$r$ values (20 or 30 seconds) reduced the token count to 13 and 12 tokens, which improved efficiency but led to a slight decline in performance, likely due to insufficient capture of finer temporal details.
%\vspace{-1.em}
\subsection{Baselines}
We demonstrate the performance of \textit{Biaxialformer} by comparing it with the following baseline methods: 1) Three \textit{CI} and \textit{CD}  Transformer-based modeling \cite{nie2022time,liu2023itransformer,zhang2023crossformer}; 2) Two Transformer-based models for EEG decoding \cite{song2022eeg,kim2024towards}. 
%\vspace{-1.em}
\begin{table}[h]
\centering
\caption{\label{tab:baselines} Results of baseline-studies on hospital E}
\begin{tabular}{|l|l|l|l|l|}
\hline
\rowcolor{lightgray} Method       & \begin{tabular}[c]{@{}l@{}}SM\end{tabular} & F1 &  AUROC  & \begin{tabular}[c]{@{}l@{}}AUPRC \end{tabular}  \\ \hline
PatchTST\cite{nie2022time}    & 0.2363                                                       &  0.5365     &  0.6210                                                          & 0.7730    \\ \hline
iTransformer\cite{liu2023itransformer} & 0.2935                                                &   0.5621    &   0.7210    &   0.7874                                                            \\ \hline
CrossFormer\cite{zhang2023crossformer}  &  0.5460                                                         &  0.6784     &   0.7878    &     0.8147                    \\ \hline
EEG-Conformer\cite{song2022eeg}  &  0.4456                                                         &  0.6372     &    0.7145   &   0.7489                                                          \\ \hline
DFformer\cite{kim2024towards}     & 0.6142                                                  &   0.7154    & 0.8383      &      0.8876                                                    \\ \hline

\textbf{Biaxialformer}   & \textbf{0.7798}                                                          & \textbf{0.7660}        &   \textbf{0.8490}   &      \textbf{0.9440}                                                    \\ \hline
\end{tabular}
\end{table}
%\vspace{-1.em}
This curated and comprehensive selection in Table \ref{tab:baselines} aims to capture the diverse landscape of contemporary \textit{CI} and \textit{CD} Transformer modeling for MTS analysis in general and multi-channel EEG analysis. \textit{Biaxialformer's} two-stage attention mechanism outperformed baseline models such as PatchTST, iTransformer, and Crossformer across multiple evaluation metrics, demonstrating its strength in modeling both intra- and inter-channel EEG features. PatchTST's \textit{CI}, while effective for intra-channel analysis, led to a failure in capturing inter-channel correlations, resulting in lower prediction accuracy. iTransformer and Crossformer, while leveraging inter-channel dependencies, struggled to capture localized temporal patterns as effectively as \textit{Biaxialformer}. The token fusion module in \textit{Biaxialformer}, with its adaptive receptive fields, provided an ideal balance between short-term detail and long-range temporal dependencies, allowing for superior generalization across tasks. It is to be noted that DFformer shares conceptual similarities with our proposed \textit{Biaxialformer}. However, DFformer still suffers from channel mixing, as in the second stage (inter-channel block) of MHA is sequentially fed with the previous, first stage MHA.
 \textit{Limitations}: While \textit{Biaxialformer} consistently outperformed the baselines, it is not without limitations. The trade-off between capturing localized and global features remains challenging, and future iterations could explore more dynamic tokenization techniques. Additionally, while iTransformer and Crossformer focus heavily on long-range dependencies, future work may further investigate how these aspects can be integrated more effectively with our framework to enhance long-term temporal modeling.
\vspace{-1.25em}
\section{Conclusion}
In this study, we introduced \textit{Biaxialformer}, a two-stage MHA Transformer-based model designed to address both intra-channel and inter-channel feature extraction in EEG signals. \textit{Biaxialformer} effectively mitigates inter-channel forgetting, a common issue in traditional CI strategies, while avoiding the channel mixing seen in models that focus too heavily on inter-channel correlations. Biaxialformer demonstrated superior performance over baseline models such as PatchTST, iTransformer, and Crossformer, particularly in predicting neurological outcomes in comatose patients. The success of the temporal-spatial fusion block, particularly when conditioned on spatial dependencies, underscores the importance of balancing both local and global relationships within EEG data. %While \textit{Biaxialformer} significantly enhances EEG decoding, challenges such as computational complexity and variability across data centers remain, opening avenues for further optimization.
\bibliographystyle{unsrt}

%\begin{multicols}{2} % Adjust the number of columns
   { \footnotesize
    \bibliography{Reference}}
%\end{multicols}

\vfill
\end{document}